\pdfoutput=1
\documentclass[a4paper,twoside]{article}

\usepackage{todonotes}
\usepackage{graphicx}
\usepackage{epsfig}
\usepackage{subcaption}
\usepackage{calc}
\usepackage{amssymb}
\usepackage{amstext}
\usepackage{amsmath}
\usepackage{amsthm}
\usepackage{multicol}
\usepackage{pslatex}
\usepackage{apalike}
\usepackage{SCITEPRESS}
\usepackage{algorithm}
\usepackage{algorithmic}
\usepackage{url}
\usepackage{bm}
\usepackage[mathscr]{euscript}

\begin{document}
 
\title{A Trend-following Trading Indicator on Homomorphically Encrypted Data}

\author{\authorname{Haotian Weng\orcidAuthor{0000-0002-3993-7621} and Artem Lenskiy\orcidAuthor{0000-0002-4745-6756}}
\affiliation{Research School of Computer Science, The Australian National University, Canberra, Australia}
\email{\{u6254332, artem.lenskiy\}@anu.edu.au}}


\abstract{Algorithmic trading has dominated the area of quantitative finance for already over a decade. The decisions are made without human intervention using the data provided by brokerage firms and exchanges.  An emerging intermediate layer of financial players that are placed in between a broker and algorithmic traders has recently been introduced. The role of this layer is to aggregate market decisions from the algorithmic traders and send a final market order to a broker. In return, the quantitative analysts receive incentives proportional to the correctness of their predictions. In such a setup, the intermediate player --- an aggregator --- does not provide the market data in plaintext but encrypts it. Encrypting market data prevents quantitative analysts from trading on their own, as well as keeps valuable financial data private. This paper proposes an implementation of a popular trend-following indicator with two different homomorphic encryption libraries --- SEAL and HEAAN --- and compares it to the trading indicator implemented for plaintext. Then, an attempt to implement a trading strategy is presented and analysed. The trading indicator implemented with SEAL and HEAAN is almost identical to that implemented on the plaintext, with the percentage error of 0.14916\% and 0.00020\% respectively. Despite many limitations that homomorphic encryption imposes on this algorithm's implementation, quantitative finance has a potential of benefiting from the methods of homomorphic encryption.}

\onecolumn \maketitle \normalsize \setcounter{footnote}{0}

\section{\uppercase{Introduction}}
\label{sec:introduction}
\noindent The most prominent approach that provides the means for data analysis and also keeps the data private is based on homomorphic encryption (HE). The idea of HE that allows computation on encrypted data was firstly proposed in the 1970s. However, no practical implementation existed until Craig Gentry proposed one in his PhD thesis in 2009 \cite{gentry2009fully}. Modern cryptosystems are capable of performing arbitrary computation on encrypted data - ciphertexts, facilitating the implementation of various data analysis tools \cite{aslett2015review}. As an active area of research, there is a multitude of HE schemes that have been proposed and implemented as open-source libraries. CKKS \cite{cheon2017homomorphic} and BFV \cite{brakerski2014efficient} are among the most popular HE schemes. Simple Encrypted Arithmetic Library (SEAL) developed by Microsoft Cryptography Research \cite{sealcrypto} implements both BFV and CKKS scheme. Homomorphic Encryption for Arithmetic of Approximate Numbers (HEAAN) is the original name of CKKS scheme developed by the Seoul National University CryptoLab which only supports CKKS scheme. A particular area of HE that has not receive well-deserved attention is the privacy-preserving analysis of time-dependent data. Such topic attracts financial industry's attention where time-series analysis is widely applied.

The financial industry is known to be extremely cautious about privacy aspects of storing, analysing and distributing data. The solutions for secure data storage have been available for decades and are already provided by numerous cloud services. Secure data distribution is nowadays an integral part of the Internet, thanks to Secure Sockets Layer (SSL). However, data analysis that preserves privacy is still in its infancy, even though, as mentioned, it has been actively developed, and adopted by the financial industry. 

Numer.ai is a hedge fund powered by thousands of independent quantitative analysts striving to outperform the market \cite{numerai}. The quantitative analysts compete with each other, and those with accurate predictions are rewarded. Numer.ai does not provide the market data in plaintext but transforms it in a form that makes it impossible to know what financial asset a particular time series represents. This, in turn, prevents quantitative analysts from trading on their own and keeps valuable financial data private. To the best of our knowledge, Numer.ai relies on proprietary obfuscation methods. 

In our scenario, we distinguish three independent players: a broker, a decision aggregator and algorithmic traders. A broker provides access to the exchanges that is a mediator between sellers and buyers. In particular, it provides such market data as order books, recent trades and price quotes, so the market participants are able to draw a trading decision. The decision aggregator (analogous to Numer.ai) receives market data $M$ in plaintext and returns market orders $O$. The data received by the aggregator is encrypted as $c_k=\mathscr{E}_1\{m_k\}$ and is sent to an algorithmic trader $T_k$. Algorithmic traders operate over the encrypted data $c_k$ and the decision $o^*_k$ drawn by the traders are also encrypted and unknown to the traders (fig. 1). These decisions are then decrypted by the aggregator $o_k = \mathcal{D}a_1\{o^*_k\}$ and transmitted to the broker in the form of market orders.

In this paper, we focus on making a very first step in the direction of applying HE in developing an algorithm that operates on encrypted time-dependent data. On the example of a popular trend following trading strategy based on Moving Average Convergence Divergence (MACD) indicator, we demonstrate how an algorithmic trader could employ methods of homomorphic encryption to make trading decisions. \footnote{\url{https://github.com/woonhulktin/HETSA}}

\begin{figure*}[!ht]
\centering
\includegraphics[clip,trim={.0cm 0cm .0cm 0cm},width=\textwidth]{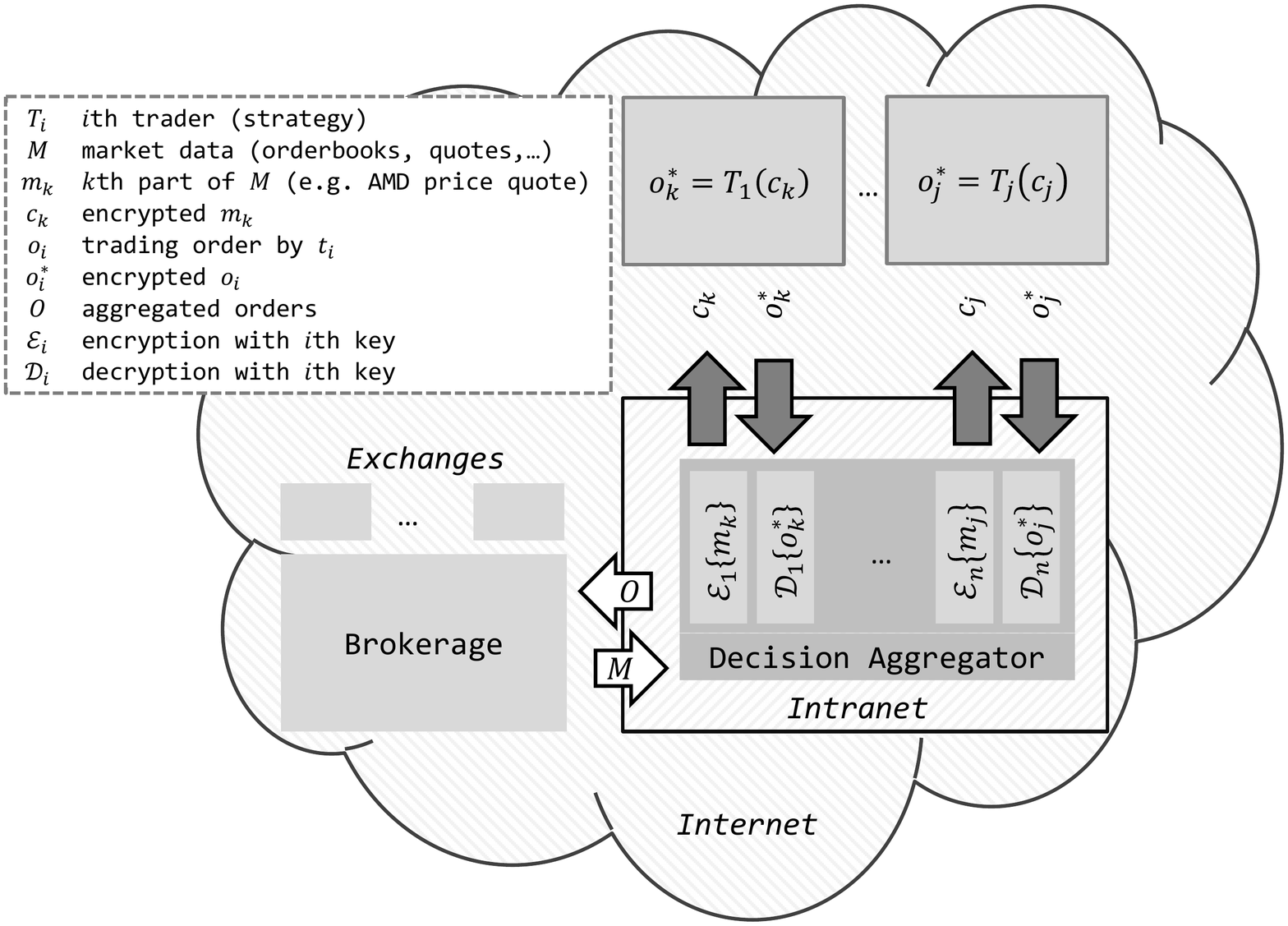}
 \caption{An intermediate layer collects encrypted decision, decrypts them and makes a final decision on whether to buy or to sell}
\label{fig:system}
\end{figure*}

\section{\uppercase{Background}}
\label{sec:motivation}

\subsection{Moving Average Convergence Divergence}
\noindent MACD is a momentum indicator which uses the difference between fast and slow moving averages to indicate market trend \cite{appel1979moving}. MACD was a valuable tool for traders during the 1980s. In this paper, we implement a modified MACD algorithm that operates on encrypted stock price.

\subsection{Homomorphic Encryption}
\noindent In the past decade, several homomorphic encryption schemes have been introduced. One of the most popular schemes is CKKS that implements approximate arithmetic of complex numbers. The scheme supports addition, subtraction and multiplication \cite{cheon2017homomorphic}.

Every operation in CKKS, especially multiplication, adds a certain amount of noise, which limits the number of operations allowed before the accumulated noise grows to the point making the final result inaccurate. Levelled schemes limit the maximal quantity of sequential homomorphic multiplications before the noise becomes intolerable \cite{brakerski2014efficient}.

To fight the noise, it introduces rescaling as well as bootstrapping. The rescaling is a scale-invariant technique that scales down the size of ciphertext modulus to reduce the noise and preserve the precision \cite{cheon2017homomorphic}. The bootstrapping operation in theory eliminates the noise accumulated throughout homomorphic computations by refreshing the noise in a ciphertext. The significant disadvantages of the bootstrapping method are the dramatic increase in computational time and significant memory consumption \cite{cheon2018bootstrapping}.

\section{\uppercase{Related Work}}
\noindent Previous studies have shown the practicality of privacy-preserving analysis of time-series data. An additive homomorphic encryption scheme was proposed to aggregate time-series data without sacrificing privacy \cite{shi2011privacy}. The ciphertexts are encrypted under different users’ secret keys respectively to achieve secure multi-party computation. Additionally, Paillier encryption scheme as a partial homomorphic encryption was applied to privacy-preserving similarity evaluation of time-series data \cite{zhu2014privacy}. Pallier scheme is adequate for computing the square of Euclidean distance as it supports homomorphic addition of ciphertexts and homomorphic multiplication to plaintexts. Another partially homomorphic-encryption-based access control construction (HEAC) was introduced to support both access control and aggregation-based computations on encrypted data \cite{burkhalter2020timecrypt}. However, these studies rely on additive homomorphic encryption schemes and thus the types of computations allowed are limited. Our approach focuses on financial time-series data and applies both homomorphic addition and homomorphic multiplication of ciphertexts to generate the trading decisions.

\section{\uppercase{Methods}}

\subsection{Data Encoding}

\noindent Both SEAL and HEAAN libraries process data in batches. This, in turn, speeds up the processing since all components of the vector are processed simultaneously \cite{chen2017simple}. However, in our scenario, the server sends encrypted price one at a time, and the remaining components are padded with zeros. This process repeats every time a new price quote is available. Therefore, only the first slot of the plaintext vector is used to store the asset price. This results in higher memory requirement and slows computation.

Once receiving the latest encrypted price quote, the algorithm appends it to the vector of previously received ciphertexts, and a MACD signal value is produced, that is either used to generate a trading decision or returned on its own.

As the share prices are rational numbers and BFV is slower than CKKS when performing multiplication followed by rescaling \cite{chen2019efficient}, the fractional encoder is chosen over the integer one. 

To test our algorithm on encrypted data, we selected Apple's daily stock price (NASDAQ: AAPL) on the interval from 06/01/2015 to 21/10/2015.

\subsection{Weighted Moving Average}

\noindent A moving average filter is a low pass filter with a linear phase shift. In the context of financial time-series, it is used to determine a trend of an asset price and is the foundation of the MACD indicator and corresponding trading strategy. The emphasis is put on recent price quotes by assigning different weighting factors to the asset prices. The original MACD indicator employs the exponential moving average (EMA), and as a first-order autoregressive filter, it requires recursion. Both SEAL and HEAAN libraries are limited by the noise that accumulates with every arithmetic operation. Without an efficient bootstrapping method, EMA is infeasible as it requires a significant number of multiplicative depths (in theory infinitely many). One solution is to replace the EMA by a non-recursive filter. One such filter is the weighted moving average (WMA). The WMA has a finite impulse response and does not require infinite multiplications. Instead, it limits the multiplicative depth by the order $n$ that defines the number of multiplications and is equal to the window size. The weighting coefficients of the WMA are chosen as follows:

\begin{equation}\label{weighting}
    \bm{w}[i] = \frac{2(i+1)}{n(n+1)}
\end{equation} \\

\noindent for $i \in [0,n)$, where $\bm{w}[i]$ is the weight and $n$ is the window size.

\begin{algorithm}
\caption{Weighted Moving Average (WMA)}
\label{WMA}
\begin{algorithmic}
\renewcommand{\algorithmicrequire}{\textbf{Input:}}
\renewcommand{\algorithmicensure}{\textbf{Output:}}
\REQUIRE a vector of ciphertexts $\bm{c}$, window size $n$
\ENSURE a vector of encrypted weighted moving averages $\bm{a}$
\STATE
\STATE $\bm{w}$ = zeros(n)
\FOR{$i$ in range [0, $n$)}
\STATE $\bm{w}$[i] = FHE.encode(2($i$+1)/($n$($n$+1)))
\ENDFOR
\STATE $\bm{a}$ = zeros($\bm{c}.size-n$)
\FOR{$i$ in range [0, $\bm{c}.size-n$)}
\STATE $\bm{a}$[i] = FHE.encrypt(0)
\FOR{$j$ in range [0, $n$)}
\STATE $r$ = FHE.multiplyConstant($\bm{c}_{i:i+n}[j]$, $\bm{w}[j]$)
\STATE $\bm{a}$[i] = FHE.add($\bm{a}$[i], $r$)
\ENDFOR
\ENDFOR
\RETURN $\bm{a}$
\end{algorithmic}
\end{algorithm}



\subsection{Moving Average Convergence Divergence}

\noindent One of the popular indicators for detecting a market turning point is the MACD indicator \cite{appel2003become}. The original MACD indicator computes two moving averages: the 12-period EMA and the 26-period EMA. Both EMAs are replaced by the WMAs for the reasons explained above.

The trading signals are triggered by the MACD signal line crossing the $x$-axis. First, a 12-period WMA $\bm{\alpha}$ and a 26-period WMA $\bm{\beta}$ are computed. Then, the differences between $\bm{\beta}$ and $\bm{\alpha}$ are calculated. Finally, a 9-period WMA $\bm{\gamma}$ is applied to the differences to produce the MACD signal line $\bm{m}$. Algorithm 2 illustrates a library-independent MACD implementation using homomorphic encryption method.

When the MACD signal line crosses the $x$-axis from below, it indicates a buy signal and when the signal line crosses the $x$-axis from above, it triggers a sell signal.

\begin{algorithm}
\caption{Moving Average Convergence Divergence Function (MACD)}
\label{MACD}
\begin{algorithmic}
\renewcommand{\algorithmicrequire}{\textbf{Input:}}
\renewcommand{\algorithmicensure}{\textbf{Output:}}
\REQUIRE a vector of ciphertexts of asset prices $\bm{d}$
\ENSURE a vector of ciphertexts of MACD signals $\bm{m}$
\STATE
\STATE $\bm{\alpha}$ = wma($\bm{d}$, 12)
\STATE $\bm{\beta}$ = wma($\bm{d}$, 26)
\STATE $\bm{\theta}$ = zeros($\bm{\beta}.size$)
\FOR{$i$ in range[0, $\bm{\beta}.size$)}
\STATE $\bm{\theta}[i]$ = FHE.sub($\bm{\alpha}_{14:\bm{\alpha}.size}[i]$, $\bm{\beta}[i]$)
\ENDFOR
\STATE $\bm{\gamma}$ = wma($\bm{\theta}$, 9)
\STATE $\bm{m}$ = zeros($\bm{\gamma}.size$)
\FOR{$i$ in range[0, $\bm{\gamma}.size$)}
\STATE $\bm{m}[i]$ = FHE.sub($\bm{\theta}_{9:\bm{\theta}.size}[i]$, $\bm{\gamma}[i]$)
\ENDFOR
\RETURN $\bm{m}$
\end{algorithmic}
\end{algorithm}

\subsection{Trading Decision}

\noindent Unfortunately due to the theoretical limitations of CKKS as well as other popular HE schemes, only basic arithmetic operations are provided. None of widespread used HE libraries is equipped with logic and relational operators on numbers \cite{acar2018survey}. Hence, there is no direct method to compare two values and deduce where the MACD signal line is greater than, less than, or equal to zero which in turn determines the time of buying, selling or doing nothing. Nevertheless, the decision of sell, hold, or buy could be associated with a decision function $o$ that produces either -1, 0 or 1 that correspond to a sell, hold or buy order.
To define a decision function $o$, we first define a $sign$ function that return the sign of a number. Then we define a vector of differences of adjacent MACD values $\bm{\delta}$ as $\bm{\delta}[i] = \bm{m}[i-1]-\bm{m}[i]$ with $\bm{\delta}[0] = 0$, and the product of consecutive MACD values $\bm{\pi}$ as  $\bm{\pi}[i] = \bm{m}[i-1]\bm{m}[i]$ for $i \in [1, \bm{m}.size)$ with $\bm{\pi}[0]$ = 0. Then the trading decision is defined as: 

\begin{equation}
    o_1(\bm{m}, i) = \frac{1}{2} sign(\bm{\delta}[i]) \cdot (sign(\bm{\pi}[i]) - 1)
\end{equation} \\

\noindent where $i \in [0, \bm{m}.size)$. \\

The range of the function above is $\{-1, 0, 1\}$. The downside of the function is its dependence on the $sign$ function that is not implementable using available HE operations. The first $sign$ function determines the trend change and the second is the moment of crossing $x$-axis.

\subsection{Polynomial Approximation of the ReLU Function}
\noindent  Given that HE schemes are limited to arithmetic operations, only polynomial functions can be implemented homomorphically. Additionally, due to the noise accumulation discussed earlier, there is a limitation on the order of a polynomial function. From this perspective, due to discontinuity of the $sign$ function, polynomials of a lower order do not approximate it well. A better function in terms of polynomial approximation is the ReLU function that is defined as follows:

\begin{equation}\label{relu}
r(x) = 
    \begin{cases}
        x & x > 0  \\
        0 & x \leq 0
    \end{cases}
\end{equation} \\

Then an equivalent to $o_1$ decision function could be implemented using ReLU as follows:

\begin{equation}
    o_2(\bm{m}, i) = - sign(\bm{\delta}[i] \cdot r(- \bm{\pi}[i]))
\end{equation} \\

\noindent where $i \in [0, \bm{m}.size)$. \\

We employed polynomial regression to approximate the ReLU function to estimate  the polynomial coefficients:

\begin{multline}\label{approx}
    \hat{r}(x) = -0.0001x^9 - 0.0003x^8+0.0025x^7 \\ +0.009x^6-0.0253x^5-0.0984x^4+0.0882x^3 \\ +0.5173x^2+0.4475x+0.0753
\end{multline}

Then using the $\hat{r}$ we define an approximation of the decision function as follows:

\begin{equation}
    \hat{o}_2(\bm{m}, i) = - \bm{\delta}[i] \cdot \hat{r}(-\bm{\pi}[i])
\end{equation}

\noindent where $i \in [0, \bm{m}.size)$.\\

We apply both $o_2$ and $\hat{o}_2$ on plaintext MACD signals and employ only $\hat{o}_2$ on encrypted MACD signals. The percentage errors between plaintext and ciphertext implementations of $\hat{o}_2$ are then compared and presented in Section 4.

\subsection{Multiplicative Depth}
\noindent Multiplicative depth is the maximal number of sequential homomorphic multiplications allowed, while the multiplication level represents the number of sequential multiplications performed on the ciphertext \cite{brakerski2014efficient}. Namely, a polynomial's multiplicative depth depends on its degree. Multiplicative depth is defined by the parameters of the HE library. Every time a multiplication operation is performed on a ciphertext, its multiplication level goes one level deeper. The number of sequential multiplications on the ciphertext are limited by the multiplicative depth. The most substantial part of the trading strategy in terms of multiplicative depth is the implementation of $\hat{r}$ function. A multiplicative depth of 3 is required to generate $\bm{m}$ but 7 more levels are needed to produce $\bm{\delta}$. Currently, $\hat{r}$ is of degree 8 and consumes 4 multiplication levels. The trading decision $\hat{o}_2$ is of degree 9. Therefore the maximum multiplication level of our approach is 9.




\subsection{Confidentiality}
\noindent CKKS scheme is an asymmetric cryptosystem with a public and a private keys. The public key is shared with both the decision aggregator and algorithmic traders while the private key is shared with the decision aggregator. Without knowing the private key, algorithmic traders as well as the attackers who hijack the communication are not able to decrypt the ciphertexts. Therefore the confidentiality of our method is well maintained by CKKS scheme. Moreover, every trader is equipped with a unique public key, that serves as a digital sign and prevents traders from impersonating each other.

\section{\uppercase{Results}}
\noindent To compare the errors between the WMA-based ciphertext and plaintext implementations, we employ the mean absolute percentage error defined as:

\begin{multline}\label{pe}
    e(\bm{x},\bm{y}) = \sum_{i=1}^{N} \frac{\lvert \bm{x}[i] - \bm{y}[i] \rvert}{\bm{y}[i]} \times 100\%
\end{multline} \\

\noindent where $i \in [0, \bm{x}.size)$ and $N = \bm{x}.size = \bm{y}.size$. $\bm{x}$ is either decrypted WMA, MACD signals or the trading decisions and $\bm{y}$ is the vector of corresponding plaintext WMA, MACD signals or trading decisions.

Table 2 presents the comparison results. As the errors of WMA and MACD signals between encrypted and plaintext analysis are insignificant, the WMA and MACD signals generated with SEAL and HEAAN are almost identical to those in plaintext. In terms of trading decisions, we compare trading decision functions over encrypted data with SEAL and plaintext data. The peaks of the approximated trading decisions generally correspond to the exact trading decisions, but there are errors around the peaks. Additionally, the error increases as the multiplication level gets deeper, and hence the percentage error of a trading decision function is larger than the that of MACD and much larger than that of WMA. The reason for the increasing error is the noise added by every arithmetic operation. 

\begin{table}[ht]
\caption{Errors between ciphertext and plaintext analysis.}\label{tab:error} \centering
\begin{tabular}{|c|c|}
    \hline
    Result & Percentage Error \\
    \hline
    WMA-SEAL & 0.00918\%  \\
    \hline
    WMA-HEAAN & 0.00019\%  \\
    \hline
    MACD-SEAL & 0.14916\%  \\
    \hline
    MACD-HEAAN & 0.00020\%  \\
    \hline
    Decision-SEAL & 3.19030\%  \\
    \hline
    Decision-HEAAN & 0.03794\%  \\
    \hline
\end{tabular}
\end{table}

The computation was conducted on the Intel(R) Core(R) CPU i7-7820HQ @ 2.9GHz with 16GB RAM. The task consisted of the following steps: (1) 200 AAPL share prices were encrypted, (2) MACD analysis was performed, (3) encrypted trading decisions were generated and (4) the decisions were decrypted. The time required to produce the MACD signal as well as the trading decision for a single day were measured and summarised in table 3. The trading indicator implemented with SEAL can also run on 1-second candles and HEAAN implementation can be applied to 15-seconds candles. The total time consumed per data unit is 0.99 and 12.34 seconds for SEAL and HEAAN implementations respectively. \\

\begin{table}[ht]
\caption{Performance of MACD and decision analysis.}\label{tab:performance} \centering
\begin{tabular}{|c|c|}
    \hline
    Method & Computation Time \\
    \hline
    MACD-SEAL & 0.73 sec \\
    \hline
    MACD-HEAAN & 7.085 sec  \\
    \hline
    Decision-SEAL & 0.26 sec  \\
    \hline
    Decision-HEAAN & 5.255 sec  \\
    \hline
    Total-SEAL & 0.99 sec  \\
    \hline
    Total-HEAAN & 12.34 sec  \\
    \hline
\end{tabular}
\end{table}

At this point, there are two significant limitations in our implementation. Firstly, the lack of recursion reduces the decision accuracy as the multiplicative depth is constrained by the HE parameters. Although, HEAAN supports bootstrapping, it introduces substantial noise and is also computationally intensive \cite{acar2018survey}. Therefore, we did not implement bootstrapping in our approach. Secondly, operation of bootstrapping is slow, due to the absence of logical and relational operators in the state-of-the-art HE libraries. Only approximate decisions can be implemented. However, the presented algorithm accurately implements the MACD indicator and could be applied in the real world applications. 

\section{\uppercase{Conclusions}}

\noindent We have implemented the MACD indicator on a stock price time-series. To the best of our knowledge, this is the first time HE methods are applied to financial time-series analysis. The algorithm implemented with SEAL is able to produce the trading indicator in less than a second and could be applied to 1-second candle data as well as to lower resolution data.  For the future work we plan to implement linear systems and corresponding recessive AR and MA based filters, including exponential moving average.

\bibliographystyle{apalike}
{\small
\bibliography{paper}}

\end{document}